\documentclass{ifacconf}

\usepackage{algorithm}
\usepackage{algpseudocode}
\algdef{SE}[DOWHILE]{Do}{DoWhile}{\algorithmicdo}[1]{\algorithmicwhile\ #1}%
\usepackage{flushend}
\usepackage{amsmath}
\usepackage{amssymb}
\usepackage{IEEEtrantools}
\usepackage{mathtools}
\usepackage{blox}
\usepackage{physics}
\usepackage{graphicx}      
\usepackage{natbib}        
\begin{document}
\begin{frontmatter}

\title{On the incremental form of dissipativity} 

\thanks[footnoteinfo]{The research leading to these results has received funding from 
the Cambridge Philosophical Society.}

\author[Cambridge]{Rodolphe Sepulchre}
\author[Cambridge]{Thomas Chaffey} 
\author[Cambridge]{Fulvio Forni}

\address[Cambridge]{University of Cambridge, Department of Engineering, Trumpington Street,
Cambridge CB2 1PZ, {\tt\small \{rs771,tlc37,ff286\}@cam.ac.uk}.}

\begin{abstract}                
      Following the seminal work of Zames, the input-output
theory of the 70s acknowledged that  incremental  properties (e.g. incremental gain) are
the relevant quantities to study in nonlinear feedback system analysis.
Yet, non-incremental analysis has dominated the use of dissipativity
theory in nonlinear control from the 80s. Results connecting dissipativity
theory and incremental analysis are scattered and progress
has been limited. This abstract investigates whether this theoretical gap is
of fundamental nature and considers new avenues to circumvent it.
\end{abstract}


\end{frontmatter}
\section{Introduction}

Research goes in circles. Incremental analysis (such as
 incremental passivity or  bounded incremental gain) dominated the landscape of 
feedback system analysis in the 60s and 70s (see e.g.  the landmark textbook of
\citet{Desoer1975}).  That line of work was pioneered in the 1960 dissertation
of Zames. It is an input-output (or external) theory, and it originated from circuit theory.

In contrast, non-incremental analysis (such as  Lyapunov stability of an equilibrium state)
dominated the nonlinear control landscape in the 80s and 90s (see e.g.  the clasical
textbooks  of \citet{Khalil2002}, \citet{Isidori1995},  and \citet{VanDerSchaft1999}). 
It is a state-space (or internal) theory, and it  originated from mechanics.

The dissipativity theory of \citet{Willems1972a,Willems1972} is like 
a trait d'union between those two periods. It connects input-output
analysis and state-space analysis, circuit theory and mechanics, internal and external properties.
Yet, in its original formulation, dissipativity theory is non-incremental, that is, focuses on deviations
from of a particular equilibrium state. This state minimizes the internal storage, when no external energy is 
supplied to the system.

This paper explores the incremental form of dissipativity, which connects the incremental input-output
properties studied in the 70s to the incremental form of Lyapunov stability. Widespread interest in the latter
 only emerged in the late 90s, after  \citet{Lohmiller1998}
advocated the value of approaching stability analysis from an incremental perspective, and to 
study global stability properties as integral forms of local contraction properties. 

One would expect that the incremental form of dissipativity connects incremental external properties
to incremental Lyapunov stability in the same way as the orignal theory connects non-incremental external
properties to non-incremental Lyapunov stability. Yet, the development of incremental
dissipativity theory over the last two decades has been scattered (see e.g.
\citep{Stan2007, Forni2013, Forni2013a, vanderSchaft2013a, Verhoek2020}).
Those limited developments contrast with the growing interest in contraction
analysis, on the one hand,
and on the renewed interest in  incremental external properties, on the other. The latter is  for instance illustrated
by today's importance
of maximal monotonicity, the incremental form of passivity, or Lispchitz continuity, the incremental form of contractivity,
 in convex optimisation \citep{Ryu2021a, Bauschke2011} and machine learning
 \citep{Szegedy2013}). The gap between those internal and external concepts calls for a bridge, which is
 the topic of incremental dissipativity.
 
 After briefly reviewing the definitions and main developments of incremental analysis,  both internal and external, we provide 
 examples that point out to difficulties in bridging them with the current theory. We argue that those limitations are of
 fundamental nature and call for new and promising research avenues in nonlinear control.

\section{The incremental form of dissipativity}

Dissipativity was introduced by \citet{Willems1972a}. A state-space model with supply rate $w$
is said dissipative if there exists a nonnegative storage $S$ such that for any $(t_1,t_0) \in \mathbb{R}_2^+$,
and any input signal $u(\cdot)$, 
$$ S(x_0) + \int_{t_0}^{t_1} w(t) \ge S(x_1) $$
where $x_1$ is the state solution at time $t_1$ corresponding to the initial condition $x_0$ at time $t_0$
and the input signal $u(\cdot)$. 

The usual form of the state-space model is the set of ordinary differential equations $\dot x=f(x,u)$, $y=h(x,u)$,
and the two important examples of supply are the passivity supply $w=\bra{u}\ket{y}=u^Ty$  and the gain supply $\norm{u}^2 -
\norm{y}^2 $. 

The incremental form of dissipativity replaces states and inputs in the original definition by {\it increments}, that is {\it differences} of states and inputs: 
$$ S(\Delta x_0) + \int_{t_0}^{t_1} \Delta w(t) \ge S(\Delta x_1), $$
where $\Delta x_1$ is the difference of state solutions at time $t_1$ corresponding to a difference of  initial conditions $\Delta x_0$ at time $t_0$
and a difference of  input signals $\Delta u(\cdot)$. The incremental supply $\Delta w$ is the supply evaluated at the difference
 of inputs $\Delta u$ and difference of outputs $\Delta y$. 
   
Under smoothness assumptions, incremental dissipativity is equivalent to differential
dissipativity.  The differential form of dissipativity is obtained by letting
increments become infinitesimal, leading to the definition
$$ S(\delta x_0) + \int_{t_0}^{t_1} \delta w(t) \ge S(\delta x_1). $$
Differential solutions of the state-space model are the solutions of the (linear) variational systems determined by the linearisation of the state-space
model around solutions. 

Incremental dissipativity is considered by \citet{Stan2007} and also \citet{Pavlov2008}, as a methodology to study synchronization of nonlinear 
oscillators. See also \citep{Verhoek2020} for a more recent development. Differential
dissipativity is proposed by \citet{Forni2013} as the generalization of differential stability to open systems. It is  
 studied by \citet{Manchester2013}, and also by \citet{Forni2018} in the context of dominance
 analysis. Differential passivity is studied by \citet{vanderSchaft2013a, Forni2013a}. 

It is usual in dissipativity theory to assume that the storage is minimum at the equilibrium state $x=0$ and
that this equilibrium trajectory corresponds to the zero input $u(\cdot) \equiv 0$. In that case,  dissipativity
is the special form of incremental dissipativity when considering only increments or deviations from the equilibrium trajectory.
A generalisation of that non-incremental definition is equilibrium-independent
dissipativity \citep{Hines2011}, which again is a special
form of incremental dissipativity, with increments only considered from equilibrium trajectories.

Quadratic storage functions of the type $S(x)=x^TPx$ occupy a special place in the theory of dissipativity, as they lead to 
algorithmic solutions of many control questions via Linear Matrix Inequalities
\citep{Boyd1994}. Likewise, the majority of applications of
incremental and differential dissipativity assume quadratic storages $S(\Delta
x)=(\Delta x)^TP(\Delta x) $, where $\Delta x = x_1 - x_2$,  or $S(\delta x)=(\delta x)^TP(\delta x) $.
Beyond linear time-invariant systems, it is of course a restriction to assume that the storage is quadratic and time-invariant, and even more so to assume that the
 storage is trajectory independent.  For a time-invariant state-space model, a general form of incremental storage is a nonnegative function of two arguments $S(x_1,x_2)$. Likewise, a  general form of differential storage is a nonnegative function of two arguments $S(x,\delta x)$, 
 with the infinitesimal increment $\delta x$ an element of the tangent space at $x$. A practical question, however, is how to construct such general
 storages in applications.

\section{Internal incremental properties}

Willems considered dissipativity theory to be the Lyapunov theory of open systems. For closed systems, the dissipation inequality reduces to
$S(x_1) \le S(x_0)$, and the storage becomes the Lyapunov function used to study the stability of, say, the dynamical system $\dot x = f(x)$.
In other words, the {\it internal} property associated to dissipativity is the classical concept of Lyapunov stability. Like dissipativity, Lyapunov
stability has historically received far more attention in its non-incremental form: the focus is on deviations with respect to the equilibrium solution $x=0$. 
The 1998 paper of Lohmiller 
and Slotine is seminal in that it highlights the value of studying Lyapunov stability as an integral form of differential stability, rather than via
the conventional non-incremental Lyapunov analysis.  The authors focus on exponential stability of the linearized dynamics along any trajectory, a property that they 
name {\it contraction}. Contraction is also studied by \citet{Russo2010,Pavlov2005}. It is 
the differential form of incremental Lyapunov stability,  first studied by 
\citet{Angeli2002}.  The concept of a  differential Lyapunov function
as a Lyapunov function constructed in the tangent bundle is developed by \citet{Forni2014}. 

The study of incremental rather than non-incremental Lyapunov stability has gained popularity in the last two decades, suggesting the relevance
of incremental stability properties in all nonlinear control applications that address questions beyond the stability of
a specific equilibrium solution. It is worth  noting that Lohmiller and Slotine consider trajectory dependent Lyapunov functions (also called
contraction metrics) $S(x,t)=\delta x ^T P(t,x) \delta x$. Likewise, Angeli considers incremental Lyapunov functions of the general form $V(x_1,x_2)$,
and Forni and Sepulchre consider trajectory dependent differential Lyapunov functions (also called Finsler metrics) of the general form $V(x,\delta x)$. 
Yet, the majority of applications only use trajectory independent quadratic Lyapunov functions.

\section{External incremental properties}

Desoer and Vidyasagar make a comprehensive study of the external properties of
nonlinear systems in their classic treatise, \emph{Feedback Systems: Input/Output
Properties} \citep{Desoer1975}.  This line of research originated in the early work
of \citet{Zames1960, Zames1966, Sandberg1964} and others.  

Define the truncation operator $P_T$ by 
\begin{IEEEeqnarray*}{rCl}
P_T u(t) = \begin{cases}
        u(t)& t < T\\
        0 & \text{otherwise}.
\end{cases}.
\end{IEEEeqnarray*}
Given a Hilbert space $\mathcal{L}$, the \emph{extension of $\mathcal{L}$}, denoted
$\mathcal{L}_e$, is the space of signals $u$ such that $P_T u \in \mathcal{L}$ for
all $T$.  The incremental gain of an operator $H$ on $\mathcal{L}_e$ is
\begin{IEEEeqnarray*}{rCl}
\sup_{u_1, u_2 \in \mathcal{L}_e, u_1 \neq u_2, T>0} \frac{\norm{P_T H u_1 - P_T
                H u_2}}{\norm{P_T u_1 - P_T u_2}}.
\end{IEEEeqnarray*}
An operator $H$ on $\mathcal{L}_e$ is said to be incrementally passive if, for all
$u_1, u_2 \in \mathcal{L}_e$, $T > 0$,
\begin{IEEEeqnarray*}{rCl}
\bra{P_T u_1 - P_T u_2}\ket{P_T H u_1 - P_T H u_2} \geq 0.
\end{IEEEeqnarray*}
These two properties are equivalent under the scattering transform, which originates
in electrical circuit theory.

The equivalent properties for operators on a Hilbert space $\mathcal{L}$ are the Lipschitz constant,
defined as 
\begin{IEEEeqnarray*}{rCl}
L = \sup_{u_1, u_2 \in \mathcal{L}, u_1 \neq u_2} \frac{\norm{ H u_1 - 
                H u_2}}{\norm{u_1 - u_2}},
\end{IEEEeqnarray*}
and incremental positivity, or monotonicity, defined as
\begin{IEEEeqnarray*}{rCl}
\bra{u_1 - u_2}\ket{H u_1 - H u_2} \geq 0.
\end{IEEEeqnarray*}
 for all
$u_1, u_2 \in \mathcal{L}$.

These two properties are fundamental in the theory of convex optimization, and indeed
the property of monotonicity forms the basis of an ever expanding literature on algorithms for
large scale and nonsmooth problems \citep{Bertsekas2011, Bauschke2011, Ryu2021a,
Ryu2016, Parikh2013}.

\section{Nonlinear circuits}

Passivity is the backbone of linear circuit theory, originating in questions about
the synthesis of dynamic behaviors in one-port circuits \citep{Cauer1926,
Brune1931, Foster1924}.  A fundamental result of circuit theory is that any passive
LTI system can be realised by interconnecting elementary circuit elements (resistors,
capacitors, inductors, and transformers).

To an extent, this key connection between circuit theory and control theory
has been lost in the nonlinear theory of passivity. It can be argued that this is the result of
focusing on the non-incremental form of passivity. 

Chua realised early that passive nonlinear circuit elements could generate rich nonlinear
behaviors when connected to batteries. For instance, a nonlinear resistor is passive
provided that its voltage-current characteristic lies in the first and third quadrant (making
the instantaneous power always non-negative), but this does not prevent the device to
have {\it negative} resistance when linearised away from zero. Such negative resistance
devices are key elements of  bistable memories, nonlinear oscillators, chaotic circuits,
and spiking circuits. They lose however a key property of the linear theory of passivity, which
is that the inverse of a passive transfer function is passive. As a consequence, the  non-incremental 
form of passivity is of little use in nonlinear circuit analysis beyond the stability analysis of the zero equilibrium.
This limitation perhaps explains why nonlinear circuits  are marginal with respect to mechanical systems
in the nonlinear textbooks of the 80s and 90s.

The situation is different with the concept of incremental passivity, or maximal monotonicity.
This concept precisely originated from attempts to extend the algorithmic tractability of passive linear circuit theory to networks of
nonlinear resistors \citep{Duffin1946, Minty1960, Minty1961, Minty1961a, Desoer1974}.
Consistent
with their focus on non-incremental analysis, monotonicity is absent from
the nonlinear textbooks of the 80s and 90s.

Our recent work \citep{Chaffey2021a}
argues that as the incremental form of passivity, monotonicity retains many of the desirable properties of linear passivity theory.
Yet, rather surprisingly, we observe in that paper that the modern state-space theory of nonlinear circuits is built from non-monotone elements!
Chua defines circuit elements from  two fundamental laws of electricity: the derivative of charge equates current ($\dot q = i$)
and  the derivative of flux equates voltage ($\dot \phi = v$). Chua defines  four key circuit elements by adding to those two
laws one monotone relation between any two of the four variables. A monotone relation $R(i,v)=0$ defines a
resistor; $C(v,q)=0$ defines a  capacitor; $L(\phi,i)=0$ defines an inductor; and $M(q,\phi)=0$
defines a memristor. But we show in \citep{Chaffey2021a} that out of those four elements, only
the resistor defines a  monotone relationship between current and voltage. As one-port circuits, 
such nonlinear capacitors, inductors, and memristors all fail to define monotone (or incrementally passive)
voltage-current relationships. 

What is wrong? Is it monotonicity that is non-physical, or should we revise Chua's definition of circuit elements?
Our view is that the passivity properties of elementary circuit elements should be trajectory independent, which
pleads for a definition of circuit elements that are passive when linearized along arbitrary trajectories.
Several recent papers have pointed out difficulties with the energy-dissipation properties of theoretical nonlinear
circuit elements (see e.g. \cite{Jeltsema2020} for a recent perspective). 
Our view is that all difficulties have to do with the distinction between passivity and monotonicity,
that is, the non-incremental and incremental forms of passivity.

\section{More puzzling examples}

According to Chua's definition, the state-space model of a current-controlled nonlinear capacitor  is
$$ \dot q =  i, \; v = c(q) $$
We show in \citep{Chaffey2021a} that this model is non-monotone from current to voltage except for
the linear capacitor $v= Cq$. In that sense, the failure of monotonicity is {\it generic} for a nonlinear 
state-space model whose block diagram is the series interconnection of an integrator and a static monotone
nonlinearity.

One could argue that any {\it integrator} should be monotone, whether linear or nonlinear: positive increments
should always integrate positively\ldots yet, it is rather common to model the saturation of 
an integrating process with a nonlinear monotone function (say sigmoidal) applied at the integrator output.
Unfortunately, such a state-space model defines a  monotone input-output operator
only when the nonlinear function is linear.

The example that originally caught our attention to the problem exposed in this paper is the potassium current model
derived by Hodgkin and Huxley in their seminal paper about the biophysical mechanism
of action potentials \citep{Hodgkin1952}.
They propose the model
$$ \dot n = \alpha(V) n + \beta(V) (1-n), \; \; I = g_K n^4 (V-V_K),$$
where the nonlinear functions $\alpha(\cdot)$ and $\beta(\cdot) $ are monotone,
the conductance $g_K$ is a positive constant, and over a  domain where $V \ge V_K$
and $n \in [0,1]$. Again, it is not difficult to show that this state-space model does not
define a monotone operator from voltage to current \citep{vanWaarde2021}. Yet, Hodgkin and Huxley derived
their model by fitting a set of trajectories that all exhibit maximal monotonicity.

Those basic examples are intriguing. They suggest an inherent difficulty to model the
external property of incremental passivity with state-space models, even when those
state-space models only interconnect LTI passive transfer
functions with monotone static nonlinearities.

\section{Discussion}

There is no difficulty in formulating the incremental form of dissipativity, as a property
that connects an internal incremental property (Lyapunov incremental stability) and 
an external incremental property associated to a particular supply (say incremental passivity).
Yet, the examples discussed above show that the simplest nonlinear state-space models that ought
to be incrementally dissipative are not. This difficulty calls for further attention.
Perhaps it explains the somewhat incremental progress of incremental dissipativity theory
over the last decade. This is in contrast with the increasing popularity of incremental stability 
and incremental passivity in control, machine learning, and optimization. 
 
 We conclude this paper by indicating two possible avenues of research 
 to bridge internal and external incremental properties.
 
 A first avenue is to broaden the internal representations of input-output operators
 beyond conventional state-space models. The recent work \citep{vanWaarde2021}
 explores a data-driven construction of kernel-based operators with desirable
 incremental external properties. In that approach, the dissipativity property is
 specified by incremental Integral Quadratic Constraints. The dissipativity
 property regularizes the data fitting using the most natural concept of
 regularised least squares, bypassing the construction of a state-space model. 
 If one adopts the viewpoint that the state-space model is a mere 
 {\it construct} for algorithmic purposes, one could argue that it can be
 dispensed with. This is in line with recent developments of data-driven control
 or system-level synthesis \citep{Wang2019a}. Finite impulse response representations
of input-output operators offer an alternative to ordinary differential equations.
 
 A second avenue is to retain the conventional state-space models
 but to restrict the external signal spaces. Classical input-output theory 
 has focused on $L_p$ signal spaces, but we know today that those spaces
 are too general for algorithmic purposes. It seems
 plausible that all the examples discussed in the present paper become monotone in suitable
 Reproducing Kernel Hilbert spaces. A well-known example of such space
 is the space of band-limited signals. Such restrictions of the signal
 space would  acknowledge that no physical model
 is meant to cover an infinite range of amplitudes and frequencies. 
 We control theorists take this limitation for granted
 when dealing with linear models but will perhaps need to
 acknowledge it more prominently when considering
 nonlinear models. It is once again striking that Zames 1960 dissertation discusses
 the importance of band-limited signals at length, twenty years
 ahead of wavelet theory and at a time when the theory
 of reproducing kernel hilbert spaces was in its infancy. Research goes in circles.

\bibliography{references}
\end{document}